\documentclass[english]{article}
\usepackage[T1]{fontenc}
\usepackage[latin9]{inputenc}
\usepackage{amsmath}
\usepackage{graphicx}
\usepackage{setspace}

\usepackage{babel}

\begin{document}

\title{Local and global persistence exponents of two quenched continuous
lattice spin models}

\date{Shyamal Bhar%
\thanks{sbhar@research.jdvu.ac.in%
}, Subhrajit Dutta%
\thanks{subhrojuphys@gmail.com%
}~$^{+}$ and Soumen Kumar Roy%
\thanks{skroy@phys.jdvu.ac.in%
}}

\maketitle
\begin{doublespace}
Department of Physics, Jadavpur University, Kolkata - 700032, India
;\newline ~~~~~$^{+}$~~Department of Physics, Chandrakona
Vidyasagar Mahavidyalaya,
\end{doublespace}

Midnapore (west) - 721201, India. 
\begin{abstract}
\begin{doublespace}
Local and global persistence exponents associated with zero temperature
quenched dynamics of two dimensional XY model and three dimensional
Heisenberg model have been estimated using numerical simulations.
The method of \textit{block persistence }has been used to find the
global and local exponents simultaneously (in a single simulation).
Temperature universality of both the exponents for three dimensional
Heisenberg model has been confirmed by simulating the stochastic (with
noise) version of the equation of motion. The noise amplitudes added
were small enough to retain the dynamics below criticality. In the
second part of our work we have studied scaling associated with correlated
persistence sites in the three dimensional Heisenberg model in the
later stages of the dynamics. The relevant length scale associated
with correlated persistent sites was found to behave in a manner similar
to the dynamic length scale associated with the phase ordering dynamics.\end{doublespace}

\end{abstract}

\section{Introduction }

\begin{doublespace}
~~~~~~~\textit{Persistence or Persistence probability} ($P(t)$)
is an important physical quantity in the field of non-equilibrium
statistical mechanics\cite{watson,snrev,purrev} . Persistence has
the ability to probe into the details of the history of the dynamics.
For a general stochastic process, it is defined as the probability
that any zero mean stochastic variable $X(t)$ has never changed its
sign since the starting time of the dynamics ($t=0$). In an extended
dynamical system we study the time evolution of the order parameter
field $\phi(\mathbf{x},t)$, which fluctuates both in space and time.
Probability that the local order parameter field (at a fixed space
$\mathbf{x}$) never changed its phase since $t=0$, is known as the
\textit{local (or site ) persistence probability} ($P(t)$). Similarly
we can define \textit{global persistence probability} \cite{global1,global2}
for the total value of order parameter (summed over entire space).
Spin systems (like Ising, Potts, XY, Heisenberg, spin nematic models
etc) defined on a lattice are nice examples of extended systems. Study
of local and global persistence exponents of such systems are of considerable
interest in the field of non-equilibrium statistical mechanics \cite{snrev,1dpotts,1dpotts2,roy_dutta}.
Persistence probability (local and global) decays with time following
the rule $L^{-\theta}(t)$ , where{\Large{} }$L(t)$ is the dynamical
growing length scale associated with the phase ordering process of
a quenched system \cite{brayrev}. $\theta$ is known as the persistence
exponent and hence there are two exponents, one is the local persistence
exponent ($\theta)$ and the other is the global persistence exponent
($\theta_{0})$ \cite{block1}. In lattice spin systems, local persistence
probability is simply the fraction of spins whose any one of the components
has never changed sign since t=0. Similarly, global persistence is
simply the probability that any one of the components of the total
value of order parameter has never changed sign since t=0. One must
take average of the \textit{persistence probabilities} corresponding
to all components and also over several initial configurations. Estimation
of the global exponent using numerical simulation is a very tedious
job as it requires a very large number of initial configurations (Ref.
22 of \cite{block1}).\textit{ Block persistence}, introduced by Cueille
and Sire \cite{block1}, provides a natural way to estimate simultaneously
both the global and the local persistence exponents in a single simulation.
Blocking (phenomenologically same as the blocking operation in Renormalization
Group methods \cite{skma} ) is nothing but summing the order parameters
(spins) over blocks of linear size $l$, where $l$ is much smaller
than the lattice size.\textit{ Block persistence} $P_{l}(t)$ (see
equation \ref{definition_PLT}) is the persistence probability associated
with the blocked spin variables (Details may be obtained from the
original work of Cueille and Sire\cite{block1,block2} ).\textit{
Block persistence} behaves like both global and local persistence
in different limits of time of the coarsening (phase ordering) process.
Persistence and scaling in two dimensional XY model has been studied
in details in a previous work \cite{roy_dutta}. However this work
was mainly restricted to the studies related to the local persistence
associated with the T=0 quenched two dimensional XY model. 

The present work is largely motivated by the work of Cueille and Sire\cite{block1,block2},
where the method of block persistence has been applied to two continuous
spin models, i.e. the two dimensional XY model and the three dimensional
Heisenberg model. Using block persistence method, both global and
local persistence exponents associated with the zero temperature quenching
dynamics of both the models have been estimated. Another interesting
phenomenon we have studied is the scaling associated with the correlated
persistence sites of three dimensional Heisenberg model. The correlated
persistence sites do not give rise to any new relevant length scale,
in fact it has been observed that it is same as the linear domain
length scale in the phase ordering kinetics. We explain in the next
section how the method of block persistence is useful in the determination
of both the persistence exponents in a single simulation and also
discuss its utility in the study of persistence associated with the
non-zero temperature quenching dynamics. 
\end{doublespace}

\section{The method of \textit{block persistence} }

~

\begin{doublespace}
As mentioned earlier, block persistence is the persistence probability
associated with the blocked spin variables. During the initial stages
of the dynamics, the dynamical length scale $L(t)$ is small compared
to the linear block size $(l)$ and block persistence behaves like
global persistence. But at the later stages of the dynamics, $L(t)$
is appreciably large as compared to the linear block size and the
block persistence behaves like local persistence. Thus, depending
on the value of $L(t)/l$, block persistence $P_{l}(t)$ behaves like
global or local persistence. A single scaling form for block persistence
given below, gives a very efficient way of determining local ($\theta$)
and global ($\theta_{0}$) persistence exponents simultaneously. The
scaling form for $P_{l}(t)$ associated with the $T=0$ quenched case
is given by, 

{\large \begin{equation}
P_{l}(t)=l^{-\theta_{0}}g\left(\frac{L(t)}{l}\right).\label{scalingzeroT}\end{equation}
}For $x\rightarrow\infty$ $(L(t)>>l$), $g(x)$ behaves like $x^{-\theta}$and
for $x\longrightarrow0$ $(L(t)<<l)$ it behaves like $x^{-\theta_{0}}$
(where $x=L(t)/l$). By choosing a proper value of $\theta_{0}$,
one should get good scaling. Slope of the log-log plot of $g(x)$
vs $x$ gives the global persistence exponent at the small $x$ region
while for the large $x$ region it gives an estimate of the local
exponent. 

When the system is quenched at temperature $T\,(\neq0)<T_{c}$ , then
spin components change sign both due to change in phase and thermal
excitations. Thermal excitations occur with decay rate given by $\tau\sim exp[-\bigtriangleup E/K_{B}T]$
(this is known as Arrhenius law) , where $K_{B}$ is the Boltzmann
constant and $\bigtriangleup E$ is the change in energy of the system
due to thermal flipping associated with a spin. The energy barrier
is of the order of the coupling constant appearing in the microscopic
interaction Hamiltonian. For the non-zero temperature quench the persistence
probability falls off much faster than the zero temperature case.
However if the thermal excitations could be eliminated properly, then
we should get temperature universality of persistence exponents. Derrida
proposed a scheme to eliminate thermal excitations in case of Ising
and Potts models \cite{bderidda}. He studied finite temperature persistence
exponent associated with non-conserved Ising model and Potts model.
In his method, two configurations, one completely uniform (system
A) and another completely random (system B), were simultaneously updated
(with time) using the same sequence of thermal noise (random numbers)
and the same Monte-Carlo algorithm (like Heat Bath, Glauber or Metropolis).
Flips in system A were only due to thermal fluctuations, whereas,
flips in system B were due to both thermal fluctuations and change
in phase. By eliminating simultaneous flips from system B, the temperature
independent persistence probability was estimated and this was same
as that of the persistence probability associated with the T=0 quench
of the system. Using this method Stauffer performed rigorous simulations
and found temperature universality of persistence exponent in the
case of Ising model \cite{DSTAUFFER}. Derrida's method is not suitable
for the conserved model and it is not trivial enough to be applied
directly to the continuous spin models. Moreover it suffers from the
problems of damage spreading \cite{damage}. The method of \textit{block
persistence }was introduced to overcome the difficulties associated
with Derrida's method. It eliminates the short scale thermal excitations
in a natural way and is readily applicable to the continuous spin
models. With increase in the level of blocking, thermal flippings
get eliminated gradually. However, the effect of thermal noise never
gets eliminated completely for any finite value of $l$. For the non-zero
value of the temperature of quench, the scaling relation for \textit{block
persistence} gets slightly modified and is given by \cite{block1},

\begin{equation}
P_{l}(t)=l^{-\theta_{0}}f\left(\frac{L(t)}{l}\right)exp\left(-\frac{t}{\tau(l)}\right).\label{scalingTnonzero}\end{equation}
where $\tau(l)$ is the effective decay rate of temperature flipping
which depends on the linear block size and temperature as well. 
\end{doublespace}

\section{Present work }

\begin{doublespace}
~~~~~~~~~~~~~~~~~In the first part of the present
work we have used the method of block persistence to find out the
local and global persistence exponents for the two dimensional XY
(O(2)) model and the three dimensional Heisenberg model (O(3)) quenched
at $T=0$ from $T=\infty$ (homogeneous phase). The first part of
our work also involves the study of the persistence probability associated
with the non-zero temperature quenching dynamics of three dimensional
Heisenberg model. From the renormalization group results it is known
that, there are only two fixed points, i.e. $T=0$ and $T=T_{c}$.
Quenching at any finite temperature below $T_{C}$ is physically similar
to that of zero temperature quench (same values of dynamical exponents).
However, for quenching at $T=T_{c}$ (known as the critical quench),
we get the different values of dynamical exponents. Thus, we should
get the same values of both persistent exponents if the temperature
of the quench is below critical temperature. It may be noted that
2-dimensional XY model is an exception, because here we have a continuous
series of fixed points (called the KT line) extending from the Kosterlitz-Thouless
transition temperature $T_{KT}$ to $T=0$ \cite{kt} and therefore
we have not studied finite temperature persistence in two dimensional
$XY$ model \cite{snmajumprivate}. In the present study we have numerically
simulated the stochastic equation of motion (with noise version of
the deterministic equation) to incorporate the non-zero temperature
effect in quenching dynamics of the three dimensional Heisenberg model.
However noise added was small enough to retain the dynamics below
critically. 

~~~~~~~~~~~~~~~~~In the second part of our work we
have studied the scaling associated with correlated persistence sites
of the three dimensional Heisenberg model quenched at T=0 from T=$\infty$.
In a previous work we have studied the persistence scaling associated
with correlated persistence sites of two dimensional XY model \cite{roy_dutta}.
We discuss in the next section the simulation procedure in details
and present the results obtained. General discussion and mathematical
formulations related to scaling associated with correlated persistence
sites is given in the later part of the next section. 
\end{doublespace}

\begin{doublespace}

\section{Simulation details and results }
\end{doublespace}

\begin{doublespace}
~~~~~~~~~~~~~~~~~The interaction Hamiltonian describing
both two dimensional XY model and the three dimensional Heisenberg
model is given by,

\begin{equation}
H=-\sum_{<i,j>}(\mathbf{\phi_{i},\phi_{j}}),\label{Hamiltonian}\end{equation}
where \textbf{$\phi$ }is the usual two or three dimensional spin
vectors (for XY or Heisenberg model) and $<i,j>$ represents the nearest
neighbour sites. The number of nearest neighbours is four for the
two dimensional XY model and six for three dimensional Heisenberg
model. The general equation of motion for both the models is given
by \cite{newman},

\begin{equation}
\frac{\partial\mathbf{\phi_{i}}}{\partial t}=\sum_{j}\mathbf{\phi_{j}}-\sum_{j}(\mathbf{\phi_{i},\phi_{j}})\mathbf{\phi_{i}}+A\,\,\eta,\label{eom1}\end{equation}
where the sum is taken over all nearest neighbour sites of i. $\eta$
represents a random number (uniformly distributed between -1 to +1)
and is associated with the amplitude factor A. For non zero values
of A, we get the effect of finite temperature of the quench. Non zero
values of A (stochastic equation) have been used to investigate the
temperature universality of persistence exponents for the three dimensional
Heisenberg model.

~~~~~~~~~ In our simulation, we have used a discretized version
of the above partial differential equation and have chosen a value
for the time step $\delta t=0.02$. We have extensively simulated
the discretized noise-free (or deterministic) version ($A=0$) of
above equation to find out the persistence exponents of both the models
and scaling associated with the correlated persistence sites for the
three dimensional Heisenberg model. The mathematical representation
of \textit{the block persistence probability} is given by,{\large \begin{equation}
P_{l}(t)=Probability[Sb_{i}(t^{'})\times Sb_{i}(0)>0,\,\,\,\,\forall\,\,\,\, t^{'}\, in\,[0,t]\,\,\,].\label{definition_PLT}\end{equation}
}where $Sb_{i}$ is the $i^{th}$ component of the \textit{blocked
spin} for a particular site in the blocked lattice. $\mathbf{Sb(t)}$
is obtained by summing over all spins \textbf{$\mathbf{S(t)}$} situated
at the sites of a block of linear size $l$. The system sizes we have
used are $2400\times2400$ for the XY-model and $144\times144\times144$
for the Heisenberg model. Owing to symmetry, we have taken average
over all the components of the spin (two for the XY model and three
for the three dimensional Heisenberg model). We have also taken average
over all the sites (of the blocked lattice) and several initial configurations.
For $l=1,$ $P_{l}(t)$ simply becomes the site persistence probability
$P(t)$. 

~~~~~~~~~The local persistence exponent for two dimensional
XY model using the \textit{block persistence} method was found to
be 0.305 ($\pm0.05)$, which agrees with the value we obtained in
our earlier work \cite{roy_dutta}. In figure 1 and 2 we have shown
the scaling associated with the block persistence probability (calculated
at T=0 by setting A=0) for the two dimensional XY model and the three
dimensional Heisenberg model. The global persistence exponent for
the two dimensional XY model was found to be $0.22(\pm0.01)$ and
that for the three dimensional Heisenberg model was found to be $0.13(\pm0.01)$,
while the local exponents were found to be 0.305 ($\pm0.05)$ and
0.50 ($\pm0.01)$ for the two models respectively. In the insets of
figure 1 and figure 2, we have plotted $log(P_{l}(t))$ versus $log(L(t))$
for various values of linear block size $l$ mentioned in the figure
caption. One should note that, the growth laws of $L(t)$, the dynamical
length scale, are different for two models. For two dimensional XY
model, it grows like $(t/ln(t))^{1/2}$, whereas for three dimensional
Heisenberg model, it follows the growth law $\sim t^{1/2}$.
\end{doublespace}

\begin{figure}
\includegraphics[scale=0.5]{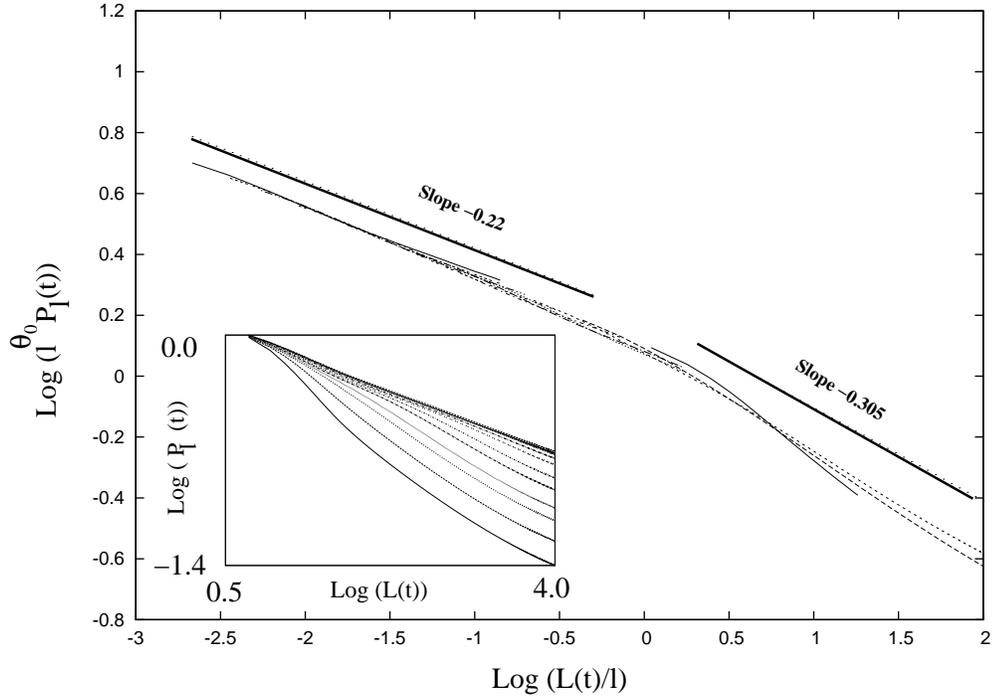}

\caption{Log -log plot of scaling function ($g(x)$ of equation 1) for block
persistence probability $p_{l}(t)$ versus $x=L(t)/l$ for $2400\times2400$
XY model at $T=0$ for linear block sizes $l=$2, 3, 4, 6, 8, 12 ,16,
20, 24 and 30. Best (to eye) scaling or data collapse was obtained
by using the value of global exponent $\theta_{0}=0.22(\pm0.01)$.
The slope of scaled function for smaller values for $x=(L(t)/l$)
is equal to the global exponent $\theta_{0}=0.22$ and is equal to
the local exponent $\theta=0.305(\pm0.05)$ for large values of $x$.
The inset shows log-log plot for block persistence probability $P_{l}(t)$
versus $L(t)$ (for the values of $l=$1, 2, 3, 4, 6, 8, 12, 16, 20,
24 and 30 from bottom to top)}

\end{figure}

\begin{figure}
\includegraphics[scale=0.5]{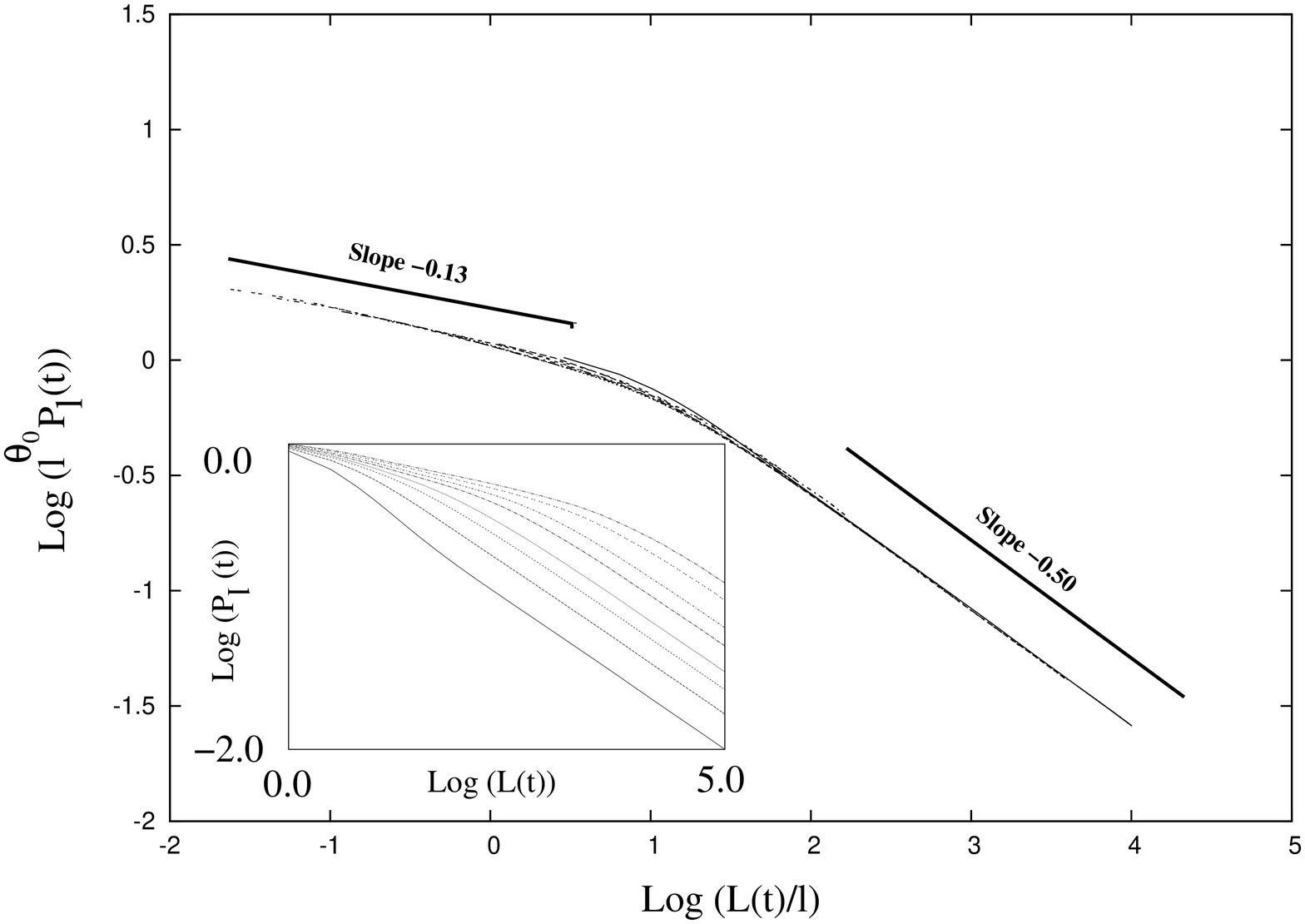}

\caption{Log-log plot of scaling function (g(x) of equation 1) for block persistence
probability $p_{l}(t)$ versus $x=L(t)/l$ for $144\times144\times144$
Heisenberg model at $T=0$ for linear block sizes $l=$2, 3, 4, 6,
8, 12 and 16. Best (to eye) scaling or data collapse was obtained
by using the value of global exponent $\theta_{0}=0.13(\pm0.01)$.
The slope of scaled function (g(x) in equation (1) ) for smaller values
for $x=(L(t)/l$), is equal to the global exponent $\theta_{0}=0.13$
and is equal to the local exponent $\theta=0.50(\pm0.01)$ , for large
values of $x$. The inset shows log-log plot for block persistence
probability $P_{l}(t)$ versus $L(t)$ (for the values of $l=$1,
2, 3, 4, 6, 8, 12 and 16 from bottom to top)}

\end{figure}

\begin{doublespace}
~~~~~~~In the figure 3 we have shown the scaling of the block
persistence probability. Scaling was good for blocks of linear size
greater than 3. Using scaling, we obtained the values of the global
and local persistence exponents which were found to be the same as
that of the T=0 case. This confirms that the persistence exponents
do not violate the temperature universality below criticality. We
have used the noise amplitude upto 0.07 and got very similar results.
In the inset of figure 3 we have shown the log-log plot of $P_{l}(t)\sim L(t)$
of the three dimensional Heisenberg model for various linear block
sizes (mentioned in the caption of the figure). The value of the noise
amplitude used was $A=0.05$. Clearly on account of non-zero temperature
effect, the decay of the blocked persistence probabilities behaves
differently from the zero temperature case for lower values of linear
block sizes $(l)$, and however the behavior of decay is found quite
similar as that of T=0 case for large values of $l$. This is because,
as mentioned earlier, the effective characteristic time ($\tau$)
in the Arrhenius law grows rapidly with $l$ and hence the effect
of temperature (see equation \ref{scalingTnonzero} ) has its role
much later---well beyond the time up to which we have performed our
simulation. Thus for larger block size, scaling is similar to the
zero temperature case and we can safely use the scaling relation for
T=0 i.e. equation \ref{scalingzeroT}. Temperature universality of
scaling function is shown in figure 4. We have shown simultaneously
the log-log plot of scaling function $g(x)=l^{\theta_{0}}P_{l}(t)$
versus $x=L(t)/l$ for the T=0 case and scaling function for non-zero
temperature multiplied by some constant factor $a_{1}$ , i.e. $g(x)=a_{1}l^{\theta_{0}}P_{l}(t)$
versus $x=a_{2}L(t)/l$, where $a_{2}$ another constant. $a_{1}$
and $a_{2}$ are same for all values of $l$. The second constant
arises due to temperature dependence of the prefactor in the growth
law for $L(t)$ \cite{brayrev}. All data shown in various figures
were obtained by averaging over 15 initial configurations (except
those mentioned in the caption). Errors in the exponents mentioned
in the text and the captions of figures for local and global exponent
were obtained by roughly estimating the region over which the collapse
appears optimal. 
\end{doublespace}

\begin{figure}
\includegraphics[scale=0.5]{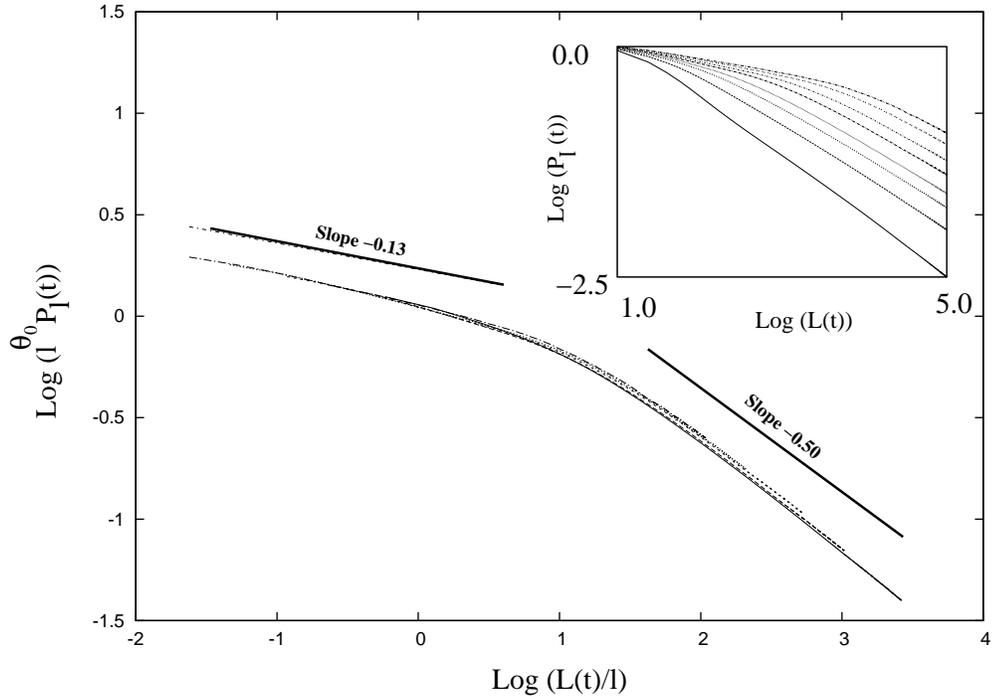}\caption{Log-log plot for scaling function of block persistence probability
$P_{l}(t)$ for $144\times144\times144$ three dimensional Heisenberg
model for noise amplitude $A=0.05$ ($l>3)$. Good collapse is obtained
with global exponent $\theta_{0}=0.13(\pm0.03)$ , which is equal
to its zero temperature value. The slope of scaled function for smaller
values for $x=(L(t)/l$) gives the global exponent $\theta_{0}=0.13$
and the local exponent $\theta=0.50$ for large values of $x$ . The
inset shows log-log plot for block persistence probability $P_{l}(t)$
versus $L(t)$ (for the values of $l=$1, 2, 3, 4, 6, 8, 12 and 16
from bottom to top )}

\end{figure}

\begin{figure}
\includegraphics[scale=0.5]{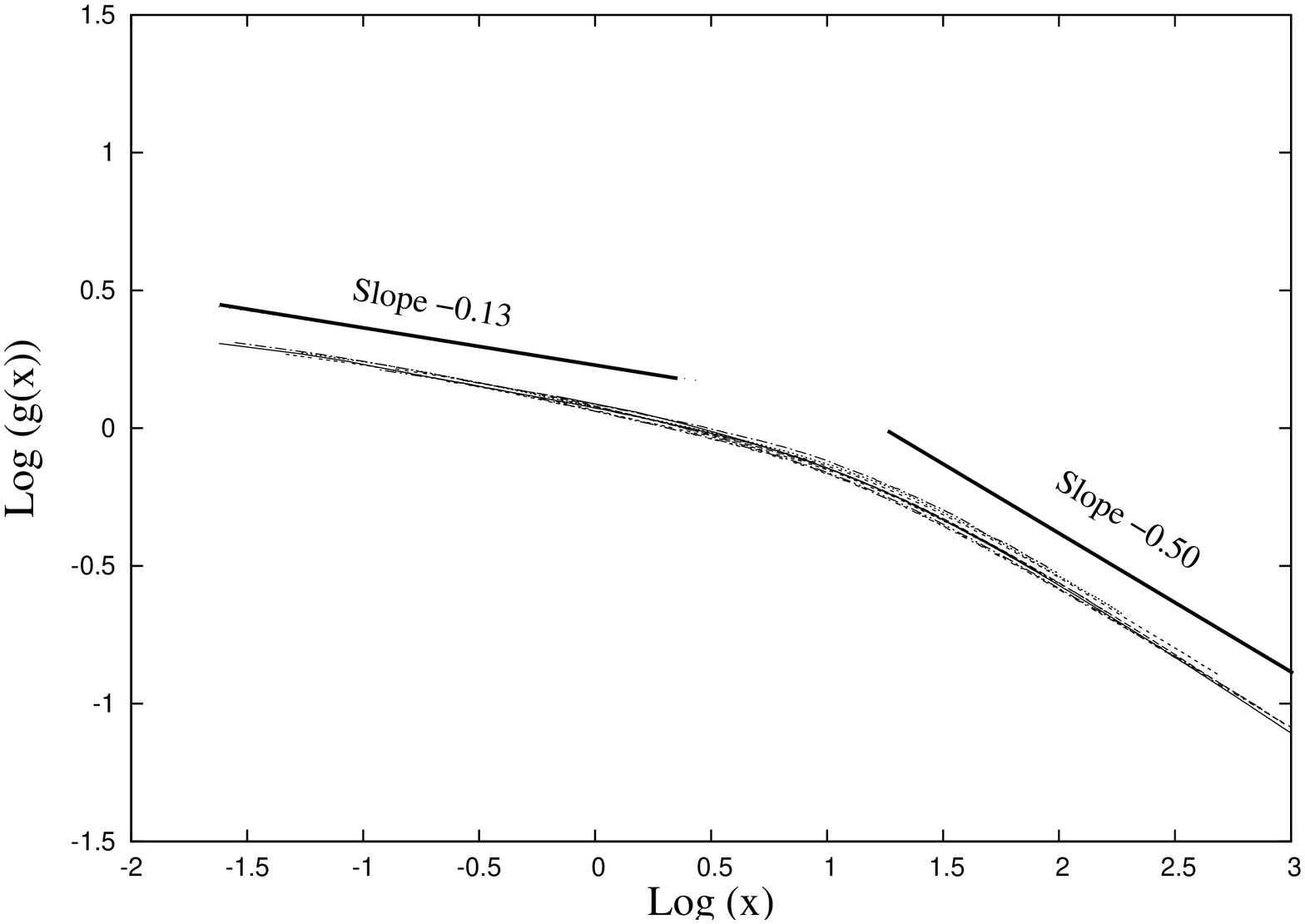}\caption{Simultaneous log-log plot of scaling function $g(x)=l^{\theta_{0}}P_{l}(t)$
versus $x=L(t)/l$ for the T=0 data and scaling function for non-zero
temperature multiplied by some constant factor $a_{1}=1.07$ ,i.e.
$g(x)=a_{1}l^{\theta_{0}}P_{l}(t)$ versus $x=a_{2}L(t)/l$, where
$a_{2}=1.02$ is another constant. Best collapse is obtained using
$\theta_{0}=0.13$. The linear block sizes used were greater than
3. }

\end{figure}

\begin{doublespace}
~~~~~~~~~ In figure 5 we have shown how the correlated regions
of persistent sites are formed at various times $t$ of the dynamics
after the quench (at T=0). Scaling and fractal formation by the correlated
persistent sites has attracted much interest and have been studied
in details by various researchers \cite{roy_dutta,manoj,bray2,jain}.
In a previous work we have studied the same for the T=0 quenched two
dimensional XY model \cite{roy_dutta}. Hence in the present work
we confine ourselves to the study of the scaling associated with the
correlated persistent sites of the three dimensional Heisenberg model.
The persistence correlation function is defined as 

\begin{align}
C(r,t) & =\frac{<\eta_{i}(t)\,\eta_{i+r}(t)>}{<\eta_{i}(t)>},\label{correlation1}\end{align}
where $<>$ represents average over several initial random configurations,
sites and components of spins. $\eta_{i}(t)=1$ if site i is persistent
at time t of the dynamics, otherwise it is 0. C(r,t) is a measure
of the probability that the $(i+r)^{th}$ site is persistent when
$i^{th}$ site is persistent. The length beyond which the persistent
sites are uncorrelated is known as persistent correlation length $(\xi(t)$).
For $r<\xi(t)$ the persistent sites are strongly correlated and the
correlation function is found to be independent of $t$ (and hence
of L(t)). In the correlated region, C(r,t) shows a power law decay
$r^{-\alpha}$ with r. For $r>>\xi(t),$ C(r,t) becomes equal to $<\eta_{i}(t)>$
or simply P(t), the site persistence probability. At $r=\xi(t)$,
continuity demands, $L^{-\theta}(t)=r^{-\alpha}$ (since $P(t)\sim L^{-\theta}(t)$).
So the persistence correlation length $\xi(t)$ should behave like
$L^{\zeta}(t)$ , where $\zeta=\theta/\alpha$. Mathematically we
write C(r,t) as,

\begin{align}
C(r,t) & \sim r^{-\alpha}\,\,\,\,\,\,\,\,\,\,\, for\,\,\,\, r<<\xi(t)\nonumber \\
 & =P(t)\,\,\,\,\,\,\,\,\,\:\, for\,\,\, r>>\xi(t)\label{correlation2}\end{align}

The scaling form for C(r,t) can be written as follows,

\begin{equation}
C(r,t)=P(t)\, f(r/\xi(t)),\label{scalingform3}\end{equation}

where $f(x)$ is given by,

\begin{eqnarray}
f(x) & \sim & x^{-\alpha}\,\,\,\, for\,\,\, x<<1\label{scalingform4}\\
 & = & 1\,\,\,\,\,\,\, for\,\,\, x>>1\nonumber \end{eqnarray}
~~~~~~~~~~~~~~~~~ In figure 6, we have shown the
variation of $C(r,t)$ with r for various values of $t$ (time after
quench) for the three dimensional Heisenberg model. From the figure
the time independence of C(r,t) for small values of r is clearly visible
(see inset). For large values of r, C(r,t) essentially becomes equal
to the persistence probability P(t). In figure 7, we have shown, the
log-log plot of scaling function of C(r,t). We obtained good collapse,
for $\zeta=1(\pm0.001)$ (and hence $\alpha=\theta)$ which implies
the persistence correlation length $\xi(t)$ behaves similar to $L(t)=t^{1/2}$.
The error in $\zeta$ mentioned here is obtained by estimating the
region over which the scaling is good to eye. The behaviour of C(r,t)
and its scaling functions are similar to that for the two dimensional
XY model \cite{roy_dutta}. 
\end{doublespace}

\begin{figure}
~~~~~~~~~~~~~~~~~~~~\includegraphics[scale=0.5]{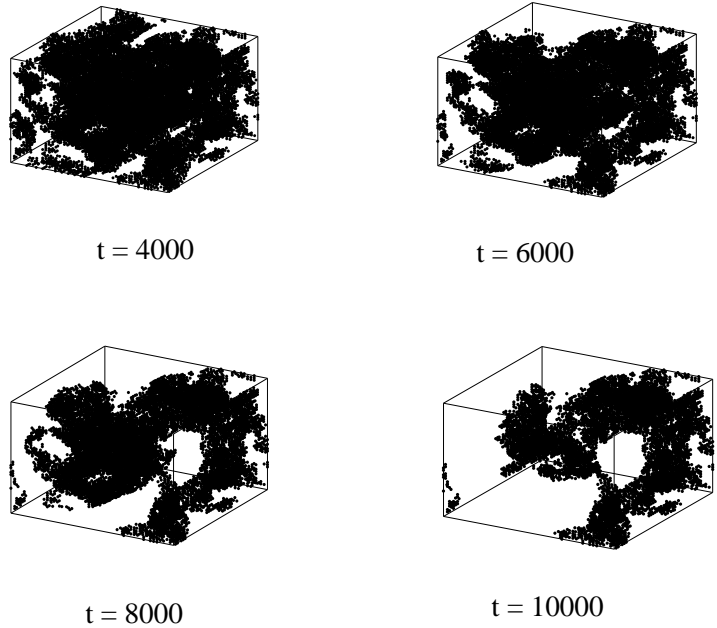}

\caption{Correlated persistent sites for $50\times50\times50$ $T=0$ quenched
Heisenberg model for time steps 4000, 6000, 8000 and 10000.}

\end{figure}

\begin{figure}
\includegraphics[scale=0.5]{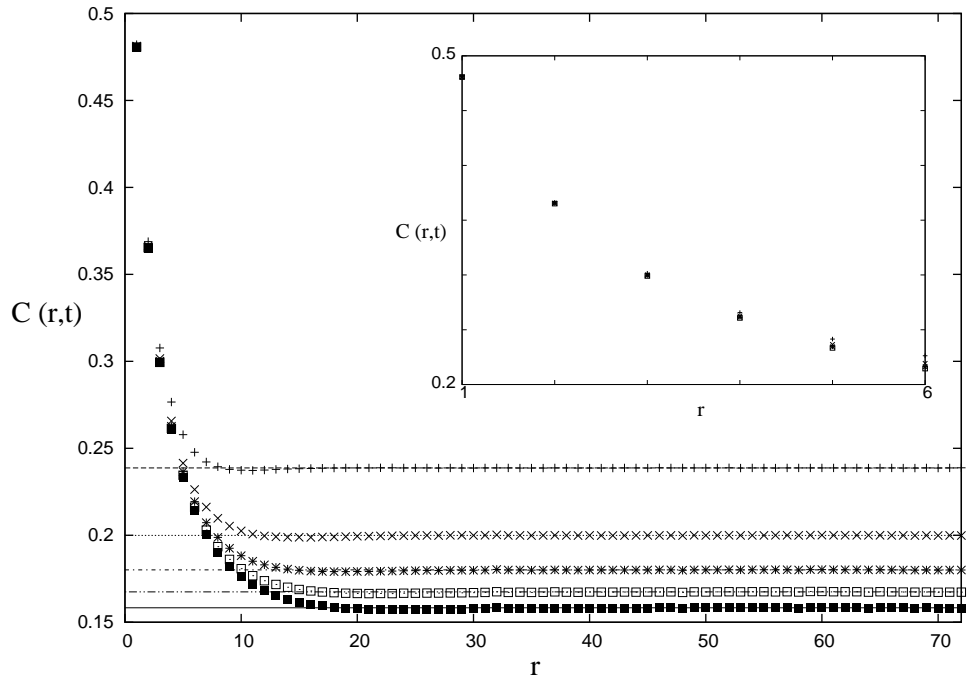}

\caption{Variation of C(r,t) with r (up to 72) for $144\times144\times144$
Heisenberg model . The data shown are for time steps t= 2000, 4000,
6000, 8000 and 10,000 (from top to bottom). For large r, C(r,t) is
same as P(t). The values of P(t) are 0.2387, 0.1999, 0.1801, 0.1675
and 0.1583 for respective time steps. The data shown are averaged
over 10 initial configurations. Inset shows the variation of C(r,t)
(at t=4000, 6000, 8000 and 10000) for small values of r. The overlapping
values of C(r,t) confirms that at the later stage of dynamics, C(r,t)
is independent of t. }

\end{figure}

\begin{figure}
\includegraphics[scale=0.5]{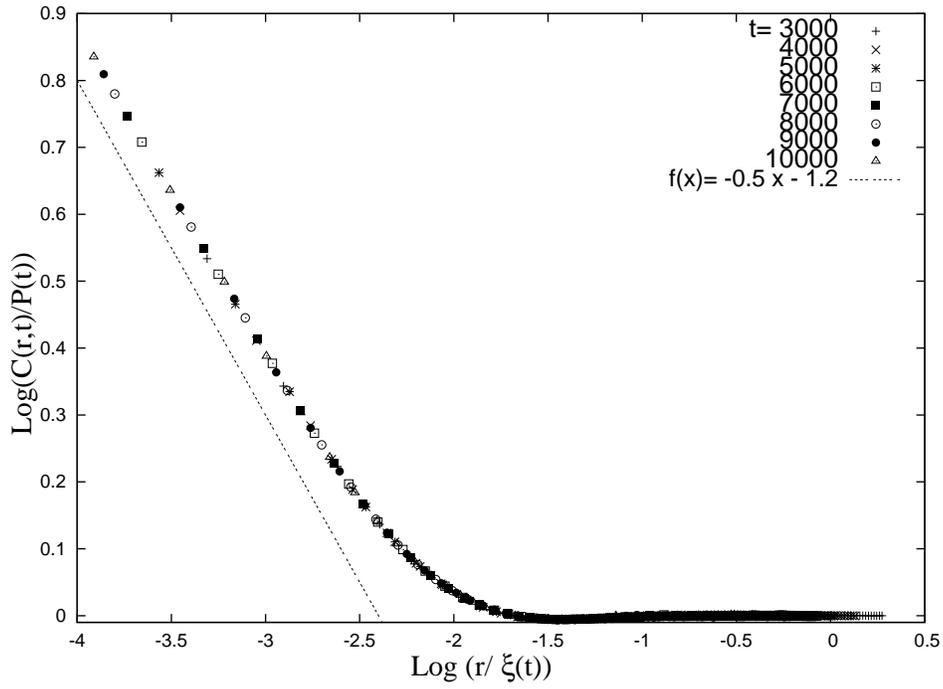}\caption{Plot of $log(C(r,t)/P(t))$ against $log(r/\xi(t))$ for time steps
t=3000, 4000, 5000, 6000, 7000, 8000, 9000 and 1000. Best collapse
was obtained for $\zeta=1\,\,(\pm0.001)$, i.e. $\xi(t)\sim L(t)\,(t^{1/2}).$
The straight line for small values of $r/\xi(t)$ has slope equal
to the local or site persistence exponent, 0.50 as expected.}

\end{figure}

\begin{doublespace}

\section{Conclusions }
\end{doublespace}

\begin{doublespace}
~~~~~~~~~~~~~~ We now sum up the findings of the present
work. Using the notion of \textit{block persistence}, we have estimated
the local and global persistence exponents for T=0 quenched two dimensional
XY model and three dimensional Heisenberg model. The local persistence
exponents were found to be equal to 0.305 ($\pm0.05)$ and 0.50 ($\pm0.01)$
for the XY and Heisenberg models respectively while the global persistence
exponents were found to $0.22(\pm0.01)$ and $0.13(\pm0.01)$. We
have found that in the case of three dimensional Heisenberg model,
the persistence exponents obey the temperature universality. Scaling
associated with the correlated persistence sites has been investigated
and we observe that the relevant length scale $\xi$ associated with
the correlated persistent sites behaves similar to $L(t)$$(\sim t^{1/2})$. 
\end{doublespace}

\section{Acknowledgements }

\begin{doublespace}
~~~~~~~~~~~~~The authors acknowledge Prof. S.N. Majumdar
for his valuable suggestions. The work was performed using the computational
resources provided by CSIR (India) (Grant No. 03 (1071)/06/EMR-II).
One of us (S.B.) acknowledges financial support from CSIR (India). 
\end{doublespace}

\break

\end{document}